\documentclass[10pt,journal]{IEEEtran}
\usepackage{graphicx}	
\usepackage[cmex10]{amsmath}
\usepackage{amssymb}
\usepackage{amsthm}
\usepackage{bm}
\usepackage{color}
\usepackage{filecontents,lipsum}
\usepackage{paralist}
\usepackage{enumitem}
\usepackage{cite}
\usepackage{longtable}

\makeatletter
\g@addto@macro\normalsize{
  \setlength\abovedisplayskip{3pt}
  \setlength\belowdisplayskip{3pt}
  \setlength\abovedisplayshortskip{3pt}
  \setlength\belowdisplayshortskip{3pt}
}
\makeatother

\makeatletter
\def\thm@space@setup{
  \thm@preskip=3pt plus 1pt minus 1pt
  \thm@postskip=3pt plus 1pt minus 1pt
}
\makeatother

\theoremstyle{plain}
\newtheorem{thm}{Theorem}

\theoremstyle{definition}

\theoremstyle{remark}
\newtheorem{rem}{Remark}

\usepackage{amsfonts}

\title{{\huge Load-Flow Solvability under Security Constraints in DC Distribution Networks}}

\author{{\normalsize Zhe~Chen,~\IEEEmembership{Student Member,~IEEE}, Cong~Wang,~\IEEEmembership{Member,~IEEE}}

\thanks{Zhe Chen and Cong Wang share equal contribution.

Zhe Chen is with Department of Electrical Engineering, Technical University of Denmark, 4000 Roskilde, Denmark (e-mail: zhech@elektro.dtu.dk).

Cong Wang is with College of Communication Engineering, Jilin University, 130022 Changchun, China (e-mail: wangcong2020@jlu.edu.cn).}}

\begin{document}
\maketitle

\begin{abstract}
We present sufficient conditions for the load-flow solvability under security constraints in DC distribution networks. In addition, we show that a load-flow solution that fulfills security constraints can be obtained via a convex optimization.
\end{abstract}

\section{Introduction}
Recently, DC distribution networks have received increasing attention. In these networks, many control procedures (i) rely on the load-flow solvability, and (ii) require that the load-flow solution (if exists) should satisfy the security constraints. However, due to non-linearity of the load-flow equations, this is hard to study. In this paper, we present sufficient conditions for the load-flow solvability under security constraints in DC distribution networks. Moreover, we show that a load-flow solution that fulfills security constraints can be obtained by solving a convex optimization.

\section{Problem Formulation}
We consider a unipolar DC distribution network that has one slack bus, {\small $N$} {\small $PQ$} buses, and a generic topology (i.e., radial or meshed). The slack bus is indexed by {\small $0$}, and the {\small $PQ$} buses are indexed by {\small $1,...,N$}. For simplicity of expression, we define
\begin{itemize}
\item {\small $\mathcal{N}=\{0,...,N\}$} as the set of all buses;
\item {\small $\mathcal{N}^{PQ}=\mathcal{N}\setminus\{0\}$} as the set of $PQ$ buses;
\item {\small $\mathcal{K}(n)\subseteq\mathcal{N}$} as the set of neighbouring buses for bus {\small $n\in\mathcal{N}$};
\item {\small $I_{\mathcal{K}(n)}(k)$} as the indicator function that has value {\small $1$} when {\small $k\in\mathcal{K}(n)$} and {\small $0$} otherwise.
\end{itemize}

Now, let {\small $v_n$}, {\small $p_n$} be the real-valued nodal voltage and nodal power injection at bus {\small $n\in\mathcal{N}$}. In addition, let positive real constant {\small $g_{nk}$} be the line conductance between buses {\small $n\in\mathcal{N}$} and {\small $k\in\mathcal{K}(n)$}. Then, the load-flow equations at {\small $PQ$} buses are formulated as follows:
\begin{small}
\begin{equation} \label{eq:load-flow}
\Big(\sum_{k\in\mathcal{K}(n)}g_{nk}\Big)v_n^2-\Big(\sum_{k\in\mathcal{K}(n)}g_{nk}v_k\Big)v_n=p_n,\quad n\in\mathcal{N}^{PQ}~.
\end{equation}
\end{small}

During power system operation, the nodal voltages and branch currents should satisfy security constraints. In the paper, we express the security constraints as follows:
\begin{small}
\begin{equation} \label{eq:security1}
v^\mathrm{min}< v_n< v^\mathrm{max},\quad n\in\mathcal{N}^{PQ},
\end{equation}
\begin{equation} \label{eq:security2}
|g_{nk}(v_n-v_k)|< i_{nk}^\mathrm{max},\quad n\in\mathcal{N}^{PQ},~k\in\mathcal{K}(n),
\end{equation}
\end{small}
where {\small $v^\mathrm{min}$}, {\small $v^\mathrm{max}$}, and {\small $i_{nk}^\mathrm{max}$} are some pre-defined positive real constants. Note that, for the development of our theoretical results, we need the security constraints to be strict inequalities.

With the above grid model and notations, we study the following problem: Given {\small $v_0$} and {\small $(p_1,...,p_N)$}, is there a unique solution {\small $(v_1,...,v_N)$} that satisfies the load-flow equations \eqref{eq:load-flow} and the security constraints \eqref{eq:security1}--\eqref{eq:security2} ?

\section{Main Results}
\subsection{Existence}
In Theorem \ref{thm:1}, we provide sufficient conditions for the existence of a solution to \eqref{eq:load-flow}--\eqref{eq:security2}.

\begin{thm} \label{thm:1}
Assume that
\begin{enumerate}
\item {\small $2v^\mathrm{min}> v^\mathrm{max}> v_0 > v^\mathrm{min}$};
\item The following convex optimization {\small $\mathrm{P1}$} is feasible.
\end{enumerate}
\begin{small}
\begin{align}
[\mathbf{P1}]\quad\max\limits_{\alpha_n,\beta_{nk},v_n}\quad&\sum_{n=1}^N v_n\notag \\
\mathrm{subject~to:}\quad&\mathrm{inequalities}~\eqref{eq:security1}-\eqref{eq:security2},\notag \\
&\quad\Big(\sum_{k\in\mathcal{K}(n)}g_{nk}\Big)\alpha_n-\sum_{k\in\mathcal{K}(n)\setminus\{0\}}g_{nk}\beta_{nk}\notag \\
\label{eq:alpha1}&=p_n+\Big(g_{n0}v_0I_{\mathcal{K}(n)}(0)\Big)v_n,\quad n\in\mathcal{N}^{PQ}, \\
\label{eq:alpha2}&v_n^2\leq \alpha_n,\quad n\in\mathcal{N}^{PQ}, \\
\label{eq:beta1}&\beta_{nk}\geq 0,\quad n\in\mathcal{N}^{PQ},~k\in\mathcal{K}(n) \\
\label{eq:beta2} &\beta_{nk}^2\leq \alpha_n\alpha_k,\quad n\in\mathcal{N}^{PQ},~k\in\mathcal{K}(n)~.
\end{align}
\end{small}
Then, for any optimizer {\small $\alpha_n^\star$}, {\small $\beta_{nk}^\star$}, {\small $v_n^\star$} of {\small $\mathrm{P1}$}, we have that {\small $(v_1^\star,...,v_N^\star)$} is a solution to \eqref{eq:load-flow}--\eqref{eq:security2}.
\end{thm}

In practice, the condition {\small $2v^\mathrm{min}> v^\mathrm{max}> v_0 > v^\mathrm{min}$} in Theorem \ref{thm:1} is normally satisfied. Consequently, for any {\small $(p_1,...,p_N)$} such that {\small $\mathrm{P1}$} is feasible, there exists at least one solution to \eqref{eq:load-flow}--\eqref{eq:security2}.

It should be noticed that, Theorem \ref{thm:1} not only gives sufficient conditions on the existence of a solution to \eqref{eq:load-flow}--\eqref{eq:security2}, but also shows that a solution can be obtained by solving the convex optimization {\small $\mathrm{P1}$}.

\begin{rem}
Consider that the numerical solvers might not be able to handle strict inequalities, we can solve the following convex optimization and verify whether the optimal nodal voltages fulfill the strict inequality \eqref{eq:security1}--\eqref{eq:security2}.
\begin{small}
\begin{align}
\max\limits_{\alpha_n,\beta_{nk},v_n}\quad&\sum_{n=1}^N v_n \notag\\
\mathrm{subject~to:}\quad&\eqref{eq:alpha1}-\eqref{eq:beta2},\notag\\
\label{eq:security3}&v^\mathrm{min}\leq v_n\leq 
v^\mathrm{max},\quad n\in\mathcal{N}^{PQ}, \\
\label{eq:security4}&|g_{nk}(v_n-v_k)|\leq i_{nk}^\mathrm{max},\quad n\in\mathcal{N}^{PQ},~k\in\mathcal{K}(n) ~.
\end{align}
\end{small}
\end{rem}

\subsection{Uniqueness}
In Theorem \ref{thm:2}, we provide sufficient conditions for the uniqueness of the solution to \eqref{eq:load-flow}--\eqref{eq:security2}.
\begin{thm} \label{thm:2}
If the following polynomial optimization {\small $\mathrm{P2}$} is infeasible, then there exists at most one solution to \eqref{eq:load-flow}--\eqref{eq:security2}.
\begin{small}
\begin{align}
[\mathbf{P2}]\quad\min\limits_{\gamma_n,v_n}\quad&\sum_{n=1}^N v_n \notag\\
\mathrm{subject~to:}\quad&\eqref{eq:security3}-\eqref{eq:security4},\notag\\
&\Bigg(2\Big(\sum_{k\in\mathcal{K}(n)}g_{nk}\Big)v_n-\sum_{k\in\mathcal{K}(n)}g_{nk}v_k\Bigg)\gamma_n \notag\\
&=\sum_{k\in\mathcal{K}(n)\setminus\{0\}}\Big(g_{nk}v_n\Big)\gamma_k,\quad n\in\mathcal{N}^{PQ}, \\
&\sum_{n=1}^N \gamma_n^2=1~.
\end{align}
\end{small}
\end{thm}
As can be seen, the infeasibility of polynomial optimization {\small $\mathrm{P2}$} is determined by only the security bounds {\small $v^\mathrm{min}$}, {\small $v^\mathrm{max}$}, {\small $i_{nk}^\mathrm{max}$} and the slack-bus voltage {\small $v_0$}.

By Theorem \ref{thm:2}, under the infeasibility of {\small $\mathrm{P2}$}, if power injection {\small $(p_1,...,p_N)$} has a solution to \eqref{eq:load-flow}--\eqref{eq:security2}, then this solution must be unique.
\begin{rem}
Due to the non-linear equality constraints, {\small $\mathrm{P2}$} is not convex. This means that the numerical solvers might not be able to directly check the infeasibility of {\small $\mathrm{P2}$}. However, we know that {\small $\mathrm{P2}$} is infeasible if any convex relaxation of {\small $\mathrm{P2}$} is infeasible. For efficiency and practicality, we suggest to take the sparsity-exploiting SDP relaxation in \cite{Lasserre3}, which has already been successfully applied to the optimal power flow problem in \cite{AppMoment1,AppMoment2,AppMoment3}.
\end{rem}

\section{Proofs}
\subsection{Proof of Theorem \ref{thm:1}}
\begin{proof}
Since {\small $(v_1^\star,...,v_N^\star)$} satisfies \eqref{eq:security1}--\eqref{eq:security2}, we only need to prove that {\small $(v_1^\star,...,v_N^\star)$} satisfies \eqref{eq:load-flow}.

First, we show {\small $\Big(v_n^\star\Big)^2=\alpha_n^\star,~\forall n\in\mathcal{N}^{PQ}$}. Suppose that {\small $\Big(v_{\tilde{n}}^\star\Big)^2<\alpha_{\tilde{n}}^\star$} for some {\small $\tilde{n}\in\mathcal{N}^{PQ}$}. As a result, we have
\begin{small}
\begin{align*}
&\Big(\sum_{k\in\mathcal{K}(\tilde{n})}g_{\tilde{n}k}\Big)\Big(v_{\tilde{n}}^\star\Big)^2-\Big(g_{\tilde{n}0}v_0I_{\mathcal{K}(\tilde{n})}(0)\Big)v_{\tilde{n}}^\star \\
<&\Big(\sum_{k\in\mathcal{K}(\tilde{n})}g_{\tilde{n}k}\Big)\alpha_{\tilde{n}}^\star-\Big(g_{\tilde{n}0}v_0I_{\mathcal{K}(\tilde{n})}(0)\Big)v_{\tilde{n}}^\star \\
=&\sum_{k\in\mathcal{K}(\tilde{n})\setminus\{0\}}g_{\tilde{n}k}\beta_{\tilde{n}k}^\star+p_{\tilde{n}}~.
\end{align*}
\end{small}
Observe that the quadratic form {\small $\Big(\sum_{k\in\mathcal{K}(\tilde{n})}g_{\tilde{n}k}\Big)\Big(v_{\tilde{n}}^\star\Big)^2-\Big(g_{\tilde{n}0}v_0I_{\mathcal{K}(\tilde{n})}(0)\Big)v_{\tilde{n}}^\star$} is convex in {\small $v_{\tilde{n}}^\star$}. Moreover, this quadratic is monotonically increasing with respect to {\small $v_{\tilde{n}}^\star$}, as
{\small $$\frac{g_{\tilde{n}0}v_0I_{\mathcal{K}(\tilde{n})}(0)}{2\sum_{k\in\mathcal{K}(\tilde{n})}g_{\tilde{n}k}}\leq\frac{v_0}{2}\leq v^\mathrm{min}< v_{\tilde{n}}^\star~.$$} Therefore, {\small $v_{\tilde{n}}^\star$} can be further increased, which contradicts the optimality of {\small $v_{\tilde{n}}^\star$}. Hence, we must have {\small $\Big(v_n^\star\Big)^2=\alpha_n^\star,~\forall n\in\mathcal{N}^{PQ}$}.

Next, we show {\small $\Big(\beta_{nk}^\star\Big)^2=\alpha_n^\star\alpha_k^\star,~\forall n\in\mathcal{N}^{PQ},~k\in\mathcal{K}(n)$}. Since {\small $\beta_{nk} \geq 0,~\forall n\in\mathcal{N}^{PQ},~k\in\mathcal{K}(n)$}, we could show {\small $\beta_{nk}^\star=v_n^\star v_k^\star,~\forall n\in\mathcal{N}^{PQ},~k\in\mathcal{K}(n)$}. Suppose that {\small $\beta_{\tilde{n}\tilde{k}}^\star<v_{\tilde{n}}^\star v_{\tilde{k}}^\star$} for some {\small $\tilde{n}\in\mathcal{N}^{PQ},~\tilde{k}\in\mathcal{K}(\tilde{n})$}. Then, we have
{\small $$\Big(\sum_{k\in\mathcal{K}(\tilde{n})}g_{\tilde{n}k}\Big)\Big(v_{\tilde{n}}^\star\Big)^2-\Big(g_{\tilde{n}0}v_0I_{\mathcal{K}(\tilde{n})}(0)+\sum_{k\in\mathcal{K}(\tilde{n})\setminus\{0\}}g_{\tilde{n}k}v_k^\star\Big)v_{\tilde{n}}^\star<p_{\tilde{n}}~.$$}
Observe that the quadratic form {\small $\Big(\sum_{k\in\mathcal{K}(\tilde{n})}g_{\tilde{n}k}\Big)\Big(v_{\tilde{n}}^\star\Big)^2-\Big(g_{\tilde{n}0}v_0I_{\mathcal{K}(\tilde{n})}(0)+\sum_{k\in\mathcal{K}(\tilde{n})\setminus\{0\}}g_{\tilde{n}k}v_k^\star\Big)v_{\tilde{n}}^\star$} is convex in {\small $v_{\tilde{n}}^\star$}. Moreover, this quadratic is monotonically increasing with respect to {\small $v_{\tilde{n}}^\star$}, as
{\small $$\frac{g_{\tilde{n}0}v_0I_{\mathcal{K}(\tilde{n})}(0)+\sum_{k\in\mathcal{K}(\tilde{n})\setminus\{0\}}g_{\tilde{n}k}v_k^\star}{2\sum_{k\in\mathcal{K}(\tilde{n})}g_{\tilde{n}k}}<\frac{v^\mathrm{max}}{2}< v^\mathrm{min}< v_{\tilde{n}}^\star~.$$}
Therefore, {\small $v_{\tilde{n}}^\star$} can be further increased, which contradicts the optimality of {\small $v_{\tilde{n}}^\star$}. Hence, we must have {\small $\beta_{nk}^\star=v_n^\star v_k^\star,~\forall n\in\mathcal{N}^{PQ},~k\in\mathcal{K}(n)$}.
\end{proof}

\subsection{Proof of Theorem \ref{thm:2}}
\begin{proof}
First, for all {\small $n\in\mathcal{N}^{PQ}$}, we have
\begin{small}
\begin{align*}
&\Bigg(2\Big(\sum_{k\in\mathcal{K}(n)}g_{nk}\Big)v_n-\sum_{k\in\mathcal{K}(n)}g_{nk}v_k\Bigg)\gamma_n-\sum_{k\in\mathcal{K}(n)\setminus\{0\}}\Big(g_{nk}v_n\Big)\gamma_k \\
=&\frac{\partial\Bigg(\Big(\sum_{k\in\mathcal{K}(n)}g_{nk}\Big)v_n^2-\Big(\sum_{k\in\mathcal{K}(n)}g_{nk}v_k\Big)v_n\Bigg)}{\partial\Big(v_1,...,v_N\Big)}
\begin{pmatrix}
\gamma_1 \\
\vdots \\
\gamma_N
\end{pmatrix}
\end{align*}
\end{small}
where {\small $\displaystyle\frac{\partial}{\partial(v_1,...,v_N)}$} is the {\small $1\times N$} Jacobian.

Let the {\small $N\times N$} Jacobian matrix of equation system \eqref{eq:load-flow} be
\begin{small}
\begin{equation*}
\mathbf{J}(v_1,...,v_N)=
\begin{pmatrix}
\frac{\partial\Bigg(\Big(\sum_{k\in\mathcal{K}(1)}g_{1k}\Big)v_1^2-\Big(\sum_{k\in\mathcal{K}(1)}g_{1k}v_k\Big)v_1\Bigg)}{\partial\Big(v_1,...,v_N\Big)} \\
\vdots \\
\frac{\partial\Bigg(\Big(\sum_{k\in\mathcal{K}(N)}g_{Nk}\Big)v_N^2-\Big(\sum_{k\in\mathcal{K}(N)}g_{Nk}v_k\Big)v_N\Bigg)}{\partial\Big(v_1,...,v_N\Big)}
\end{pmatrix}.
\end{equation*}
\end{small}
Then, it can be seen that {\small $\mathbf{J}(v_1,...,v_N)$} is non-singular in {\small $\{v_n,n\in\mathcal{N}^{PQ}:\mathrm{inequalities~\eqref{eq:security1}-\eqref{eq:security2}}\}$} whenever the optimization {\small $\mathrm{P2}$} is infeasible. 

Second, {\small $\{v_n,n\in\mathcal{N}^{PQ}:\mathrm{inequalities~\eqref{eq:security1}-\eqref{eq:security2}}\}$} is convex, since \eqref{eq:security1}--\eqref{eq:security2} are linear constraints on {\small $v_1,...,v_N$}. 

By Theorem 2 in \cite{Admissibility}, we know that there exists at most one solution to \eqref{eq:load-flow}--\eqref{eq:security2} if {\small $\mathrm{P2}$} is infeasible.
\end{proof}

\bibliographystyle{IEEEtran}

\bibliography{Paper}
\end{document}